\documentclass[aps,twocolumn,amsmath]{revtex4-1}
\usepackage{bbm}
\usepackage{mathrsfs}
\usepackage{amsmath}
\usepackage{amsfonts}
\usepackage[colorlinks=true, 
            citecolor=blue, 
            anchorcolor=blue, 
            linkcolor=blue, 
            urlcolor=blue]{hyperref}
\usepackage{graphicx,epstopdf}
\usepackage{subfigure}
\usepackage{epsfig}
\usepackage{dcolumn}
\usepackage{bm}
\usepackage{color}
\usepackage{natbib}
\usepackage{amssymb}
\usepackage{xcolor}
\usepackage{braket}
\begin{document}

\title{A Dirac-fermion approach and its application to design high Chern numbers in magnetic topological insulator multilayers}

\author{Dinghui Wang$^{1,\#}$, Huaiqiang Wang$^{1,\#}$, and Haijun Zhang$^{1,2,\ast}$}

\affiliation{
 $^1$ National Laboratory of Solid State Microstructures, School of Physics, Nanjing University, Nanjing 210093, China\\
 $^2$ Collaborative Innovation Center of Advanced Microstructures, Nanjing University, Nanjing 210093, China\\
}

\begin{abstract}
    Quantum anomalous Hall (QAH) insulators host topologically protected dissipationless chiral edge states, the number of which is determined by its Chern number. Up to now, the QAH state has been realized in a few magnetic topological insulators, but usually with a low Chern number. Here, we develop a Dirac-fermion approach which is valuable to understand and design high Chern numbers in various multilayers of layered magnetic topological insulators. Based on the Dirac-fermion approach, we demonstrate how to understand and tune high Chern numbers in ferromagentic MnBi$_{2}$Te$_{4}$ films through the van der Waals (vdW) gap modulation. Further, we also employ the Dirac-fermion approach to understand the experimentally observed high Chern numbers and topological phase transition from the Chern number $C=2$ to $C=1$ in the [3QL-(Bi,Sb)$_{1.76}$Cr$_{0.24}$Te$_{3}$]/[4QL-(Bi,Sb)$_{2}$Te$_{3}$] multilayers. Our work provides a powerful tool to design the QAH states with a high Chern number in layered magnetic topological insulator multilayers. 
\end{abstract}

\email{zhanghj@nju.edu.cn}

\maketitle

\section{Introduction\label{sec1}}

The discovery of topological insulators (TIs) \cite{HasanRMP2010,qixiaoliangRMP2011,BHZScience2006,koenig2007,zhanghaijun2009topological,YLChenScience2009} encourages researchers to study Chern insulators with a nonzero Chern number. The Chern number, also known as the Thouless-Kohmoto-Nightingale-Nijs number in condensed matter physics, serves as a topological invariant to characterize the quantum anomalous Hall (QAH) state \cite{TKNN}. In a Chern insulator, there are topologically-protected gapless chiral edge states \cite{qixiaoliangPRB2008, wenghongming2015, liuchaoxingARCMP2016}, the number of which is determined by its Chern number through the bulk-boundary correspondence. To achieve a nonzero Chern number, the time-reversal symmetry must be broken. Haldane first creatively proposed an honeycomb lattice model with a periodic magnetic flux to realize the QAH state without an external magnetic field \cite{haldane1988PRL}, though it is still hard to implement in experiments. Interestingly, the QAH effect has been realized in some recent state-of-the-art experiments, though the working temperature is still quite low (up to only a few Kelvin). In those reported experiments, the Chern number $C=1$ is widely observed \cite{liuchaoxingARCMP2016}, for example, in Cr- or V-doped (Bi,Sb)$_2$Te$_3$ films \cite{zhangjinsong2013topology,changcuizu2013experimental, kouxufeng2014scale, CheckelskyNP2014,xiaodiPRL2018, grauerPRL2017,MasatakaSciAdv2017,BestwickPRL2015}, and the recently discovered antiferromagnetic (AFM) TI MnBi$_{2}$Te$_{4}$ films \cite{liuchangNM2020,dengyujun2020quantum}. Besides, the QAH effect with the Chern number $C=1$ is also observed in twisted bilayer graphene, driven by the strong interaction \cite{serlin2020intrinsic,aaronScience2019}.

\begin{figure}[tpb]
    \includegraphics[width=0.45\textwidth]{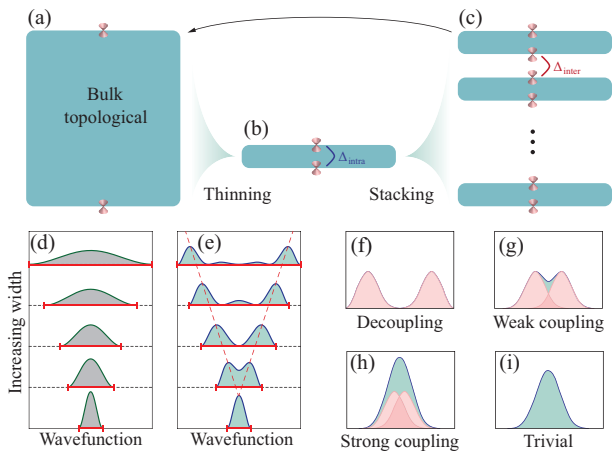}
    \caption{(a) Schematic of a 3D topological insulator (TI) with two Dirac-cone surface states on the top and bottom surfaces. (b) When the 3D TI is reduced to a thin film,  the two Dirac-cone surface states will gradually couple with each other, which is defined as a block layer (BL) with the $\Delta_{\textrm{intra}}$ denoting the intra-layer coupling. (c) Illustration of a multilayer structure constructed through stacking BLs in (b). The wave function profile from $k\cdot p$ analysis of the low energy state with reducing film thickness for (d) the topologically trivial case and (e) topologically nontrivial case. For the topologically nontrivial (trivial) case, the wave function indicates that the low energy state is a surface (bulk) state. (f,g,h) When the thickness of a TI film is gradually reduced, the Dirac-cone surface states get to couple with each other, seen in (f), (g), and (h), where the pink peaks schematically indicate the projected Dirac-cone surface states. The Dirac-cone states can be distinguished even at the ultra-thin limit (h), which is essentially different from that in the trivial case (i). }
    \label{fig1}
\end{figure}

It is well known that the chiral edge states in Chern insulators can carry dissipationless electric currents, which give rise to a quantized Hall conductivity $\sigma_{xy}=C\frac{e^2}{h}$, where the $C$ denotes the Chern number. Therefore, the Chen insulator with a high Chern number ($C>1$) is not only valuable for the fundamental topological physics, but also promising to improve the performance in next-generation electronic devices. To this end, Wang {\it{et al.}}~\cite{wangjing2013quantum} proposed the superlattices of Cr-doped Bi$_{2}$(Se$_{x}$Te$_{1-x}$)$_3$ TI for Chern insulators with $C=2$, and Fang {\it{et al.}} \cite{fangchen2014large} proposed to reach Chern numbers between $C=-4$ and $C=4$ in topological crystalline insulator thin films through the ferromagnetic (FM) doping. Jiang {\it {et al.}} \cite{jianghuaPRB2012} proposed realizing higher Chern numbers through doping magnetic elements in a multilayer of TI films. Recently, a Chern number $C=3$ was predicted in double septuple-layer (SL) MnBi$_2$Te$_4$ thin films under an external electric field \cite{dushiqiao2020berry}, and $C=N$ was predicted in $2N$-SL MnBi$_2$Te$_4$ with alternate stacking orders \cite{ZhuwenxuanPRB2022}. In experiments, a high Chern number (up to $C=5$) has been realized in [(Bi,Sb)$_{1.76}$Cr$_{0.24}$Te$_{3}$]/[(Bi,Sb)$_{2}$Te$_{3}$] multilayers based on the strategy of direct stacking \cite{zhaoyifan2020tuning}. However, Chern insulators with a high Chern number are still very rare and highly desired.

In this work, we develop an effective Dirac-fermion approach to investigate the low-energy electronic structure and design high Chern numbers in multilayers of layered magnetic TIs. In Sec.~\ref{sec2}, we first show the stable existence of local Dirac-cone surface states in TI films, even at the ultra-thin limit. Then, we build the Dirac-fermion approach to describe the low-energy electronic structures of a large variety of multilayers of layered (magnetic) TIs. In Sec.~\ref{sec3}, we demonstrate the topological phase transition with the Chern number from $C=0$ to $C=1$ in one magnetic block layer (BL), depending on the competition of the magnetic exchange field and the intra-layer coupling. We further show how the highest Chern number $C=N$ can be obtained in a film of $N$ magnetic BLs. As applications of the Dirac-fermion approach, in Sec.~\ref{sec4}, we study the FM MnBi$_2$Te$_4$ films with a high Chern number predicted. In Sec.~\ref{sec5}, we further employ the Dirac-fermion approach to understand the experimentally observed topological phase transitions in [3QL (Bi,Sb)$_{1.76}$Cr$_{0.24}$Te$_{3}$]/[4QL (Bi,Sb)$_{2}$Te$_{3}$] multilayers. Finally, a summary is given in Sec.~\ref{sec4}.

\section{Dirac-fermion approach\label{sec2}}

TIs are insulating in the bulk but metallic on the surface due to the existence of gapless Dirac-cone surface states, guaranteed by the topological invariant $Z_{2}$ \cite{FuliangZ2PRL2007, MooreZ2PRB2007, RoyZ2PRB2009}, schematically shown in Fig.~\ref{fig1}(a). Here, we take the typical layered TI Bi$_{2}$Se$_{3}$ \cite{zhanghaijun2009topological} as an example to demonstrate how to build the Dirac-fermion approach for multilayers of layered (magnetic) TIs. It is well known that Bi$_{2}$Se$_{3}$ has a large bulk band gap, in which a gapless helical spin-momentum locked Dirac-cone surface state arises, as directly observed by angle resolved photoemission spectroscopy experiments \cite{YLChenScience2009, xia_naturephysics_2009}. The low-energy effective $k\cdot p$ Hamiltonian for the helical Dirac-cone surface state to linear order in $k$ is given by \cite{Liu2010Model}
\begin{equation}
    H_D(\textbf{k})=\pm v(k_x \sigma_y-k_y \sigma_x ),
\end{equation}
where $v$ is the Fermi velocity, the Pauli matrices act on the spin subspace, and '$\pm$' denote the opposite helicities of Dirac states localized on opposite surfaces. Here, the higher-order hexagonal warping effect \cite{FuliangWarping2009} has been ignored, for we are just interested in the low-energy electronic structure around the $\Gamma$ point. Because the Dirac-cone surface state localizes on the surface layers, for example, 2-quintuple-layer (2-QL) region for Bi$_{2}$Se$_{3}$ \cite{zhang2010njp}, when a TI film is thick enough, the top and bottom surface states get completely decoupled, as schematically shown in Fig.~\ref{fig1}(a). However, along with reducing the film thickness, the top and bottom surface states will gradually couple with each other with a finite intra-layer coupling $\Delta_{\textrm{intra}}$, as schematically shown in Fig.~\ref{fig1}(b).

A natural question is whether the Dirac-cone surface states stably exist, when the TI film gets to its ultra-thin limit (e.g. 1 QL Bi$_{2}$Se$_{3}$ film)? To address this issue, we begin with the four-band bulk $k\cdot p$ model of three-dimensional Bi$_{2}$Se$_{3}$-family TIs \cite{zhanghaijun2009topological}, 
\begin{equation}
  H_{\mathrm{bulk}}(\textbf{k})=\begin{pmatrix}
  M(\textbf{k})& A_1k_z&0 &A_2k_{-} \\
  A_1k_z&-M(\textbf{k}) &A_2k_{-}&0 \\
  0&A_2k_{+}&M(\textbf{k}) &-A_1k_z\\
  A_2k_{+} & 0 &-A_1k_z& -M(\textbf{k})
\end{pmatrix}
+\epsilon_0(\textbf{k}).
\label{eq1}
\end{equation}
where $k_{\pm}=k_x\pm ik_y$, $\epsilon_0=C+D_1k_z^2+D_2k_{\parallel}^2$ and $M(\textbf{k})=M+B_1k_{z}^2+B_2k_{\parallel}^2$. The parameters of $A_{1,2},$ $C$, $D_{1,2}$, $M$ and $B_{1,2}$ can be taken from the previous work on Bi$_{2}$Se$_{3}$ \cite{zhanghaijun2009topological,Liu2010Model}, in which the curvature parameters $B_{1,2}$ have the same positive sign ($B_{1,2}$ is henceforth set to be positive). In the above $k\cdot p$ model, $M <0$ indicates an inverted band structure at $\Gamma$, and describes a Bi$_{2}$Se$_{3}$-like TI, while $M>0$ describes a topologically trivial insulator. In order to see the evolution of Dirac-cone surface states along with reducing the thickness of a TI film, we employ the method based on the quantum well states to calculate the energy spectrum and wavefunction distributions \cite{liuchaoxing2010oscillatory}.

For the topologically trivial case ($M>0$), the wave functions of the low-energy state are calculated for the films with reducing the thickness, shown in Fig.~\ref{fig1}(d). It is clear that the low-energy state is not a surface state but a bulk state, which is not surprising because the $k\cdot p$ model with $M>0$ describes a topologically trivial insulator without topologically protected surface state. Differently, for the topologically nontrivial case ($M<0$), the calculated wave functions of the low-energy state are shown in Fig.~\ref{fig1}(e). As expected, surface states are clearly seen to exist locally on the top and bottom surfaces when the film is thick enough. When reducing the film thickness\cite{zhouFiniteSizeEffects2008}, the top and bottom surface states gradually merge with a finite intra-layer coupling $\Delta_{\textrm{intra}}$. For the weak coupling, the top and bottom surface states can be visibly distinguished, as shown in Fig.~\ref{fig1}(f). When the intra-layer coupling further increases, the two surface states get to significantly merge with each other, as shown in Fig.~\ref{fig1}(g). Especially, when the intra-layer coupling is strong enough, for example, at the ultra-thin limit of 1-QL Bi$_{2}$Se$_{3}$, the wave functions seem to have one peak, but the surface states can still be decomposed (see Appendix \ref{appendix1}), as shown in Fig.~\ref{fig1}(h), which is essentially different from the single peak of wave functions in the topologically trivial case ($M>0$) in Fig.~\ref{fig1}(i). Therefore, the results indicate the stable existence of Dirac-cone surface states on both surfaces for TI thin films, as long as its bulk Hamiltonian has $M<0$, which guarantees the validity to build the Dirac-fermion approach for various multilayers of layered (magnetic) TIs. It is also worth mentioning that the previous work \cite{shanEffectiveContinuousModel2010} has strictly derived the local Dirac-cone surface states for TI films, which remains valid down to the ultra-thin limit.

Then, we show how to build a Dirac-fermion model for a multilayer of TIs. First of all, we take a TI film as one BL, which can consist of one or several unit layers. For Bi$_{2}$Se$_{3}$, the unit layer is a QL (Se-Bi-Se-Bi-Se). For each BL with two Dirac-cone surface states coming from its top and bottom surfaces, respectively, the effective low-energy Hamiltonian by the Dirac-fermion model is written as
\begin{equation}\label{eqBL}
    H_{\textrm{BL}}(\textbf{k})= v(k_x \sigma_y-k_y \sigma_x )\tau_{z}+ \Delta_{\textrm{intra}}\tau_{x},
\end{equation}
where $\tau_{x,z}$ are Pauli matrices acting in the subspaces of top and bottom surface states. Here, $\tau_z$ denotes the opposite helicity of the top and bottom Dirac-cone states, and $\Delta_{\textrm{intra}}$ denotes the intra-layer coupling of the two Dirac-cone states within each BL. Strictly speaking, apart from the constant coupling term, the $\Delta_{\textrm{intra}}$ also contains $k$-dependent terms, such as the quadratic terms derived in Refs.~\cite{Lu2010, yu2010quantized}. However, since we are mainly concerned with the low-energy physics around the $\Gamma$ point, we treat the $\Delta_{\textrm{intra}}$ as a constant term and neglect these higher-order terms. Moreover, although it has been shown that the $\Delta_{\textrm{intra}}$ may even change its sign as the BL's thickness reduces \cite{Lu2010},  it is found that its sign does not lead to essential difference for the following discussion, so the $\Delta_{\textrm{intra}}$ is set to be positive unless we specifically state.

Furthermore, TI BLs can be stacked into a multilayer, as schematically shown in Fig.~\ref{fig1}(c). Here, we consider the BLs are coupled by a weak van der Waals (vdW) interaction, which is valid for layered TIs. Naturally, the bottom Dirac-cone state of each BL couples with the top Dirac-cone state of its below adjacent BL. We take the parameter $\Delta_{\textrm{inter}}$ to denote the inter-layer coupling of these two Dirac-cone states. Therefore, the Dirac-fermion Hamiltonian of a multilayer can be directly written as, 
\begin{equation}\label{eqmultilayer}
    H(\textbf{k})=\begin{pmatrix}
    H_{\textrm{BL}}(\textbf{k})  &\tilde{\Delta}_{\textrm{inter}}   &    &    \\
    \tilde{\Delta}_{\textrm{inter}}^{\dagger} & H_{\textrm{BL}}(\textbf{k}) & \tilde{\Delta}_{\textrm{inter}} &  \\
         & \tilde{\Delta}_{\textrm{inter}}^{\dagger} & H_{\textrm{BL}} (\textbf{k})   & \ddots  \\
         &        &  \ddots   &\ddots
        \end{pmatrix}.
\end{equation}
Here, $\tilde{\Delta}_{\textrm{inter}}=\Delta_{\textrm{inter}}\tau_{-}$, with $\tau_{-}=(\tau_{x}-i\tau_{y})/2$, where $\tau_{x,y}$ are Pauli matrices acting on the top and bottom surface states for each BL.

The above Dirac-fermion approach is unified to describe the low-energy electronic structures of a wide diversity of multilayers of various layered (magnetic) TIs. It not only offers a large freedom in tuning parameters ($\Delta_{\textrm{intra}}$, $\Delta_{\textrm{inter}}$, electric field $E$, magnetic exchange field $m$, etc.), but also can often provide a simple and clear physical picture. For example, as an important regulatory means, the external electric field can be directly simulated through tuning on-site potentials in the Dirac-fermion approach, because the Dirac-cone state is local in real space. When a magnetic TI is considered, a mass term $m\sigma_z$ induced by breaking the time-reversal symmetry is directly introduced into the Dirac-cone Hamiltonian $H_D(\textbf{k})$. A notable example is MnBi$_{2}$Te$_{4}$, known as a layered A-type AFM TI, which can be regarded as a multilayer of a magnetic TI. Interestingly, its phase diagram can be investigated with the Dirac-fermion approach as we show in the following section. It is worth mentioning that, in our previous work, we have employed a three-Dirac-fermion model \cite{wangdinghui2022three} to discover a topologically-protected gapless surface state for layered magnetic TIs under the vdW modulation, which provides a unique intrinsic mechanism for understanding the puzzle of the observed gapless surface state of MnBi$_{2}$Te$_{4}$ \cite{chenyulin2019PRX, dinghong2019PRX, shikin2021PRB, liuchang2019PRX, nevola2020PRL}. Further, the MnBi$_{2}$Te$_{4}$-family materials, such as, MnBi$_{4}$Te$_{7}$ \cite{dingCrystalMagneticStructures2020, vidalTopologicalElectronicStructure2019, wuDistinctTopologicalSurface2020, yanAtypeAntiferromagneticOrder2020} and MnBi$_{6}$Te$_{10}$ \cite{shikinElectronicSpinStructure2022, shiMagneticTransportProperties2019,vidalOrbitalComplexityIntrinsic2021,yanAtypeAntiferromagneticOrder2020}, can be regarded as multilayers of magnetic TI MnBi$_{2}$Te$_{4}$ and nonmagnetic TI Bi$_{2}$Te$_{3}$, and can also be described by the Dirac-fermion approach. In addition, multilayers of TIs and topologically trivial insulators can also be well investigated through the Dirac-fermion approach, for example, the multilayer of TI (Bi,Sb)$_{2}$Te$_{3}$ and heavily Cr-doped (Bi,Sb)$_{2}$Te$_{3}$.

\section{Dirac-fermion approach for magnetic multilayers\label{sec3}}
\begin{figure*}[tbp]
    \includegraphics[width=\textwidth]{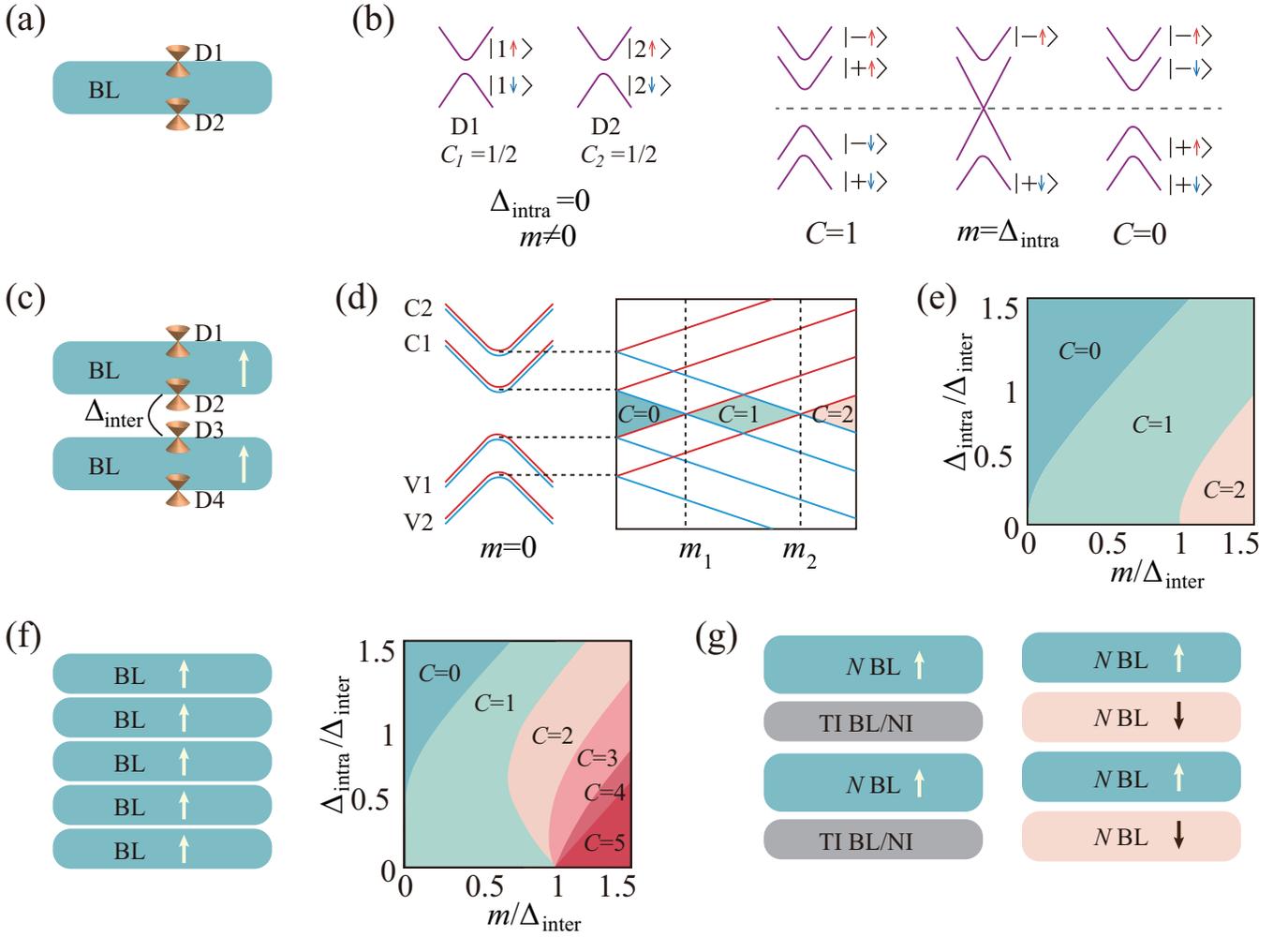}
    \caption{(a) Schematic of a BL with two local Dirac-cone states (D1 and D2). (b) Band structure from the Dirac model of a BL in the topological phase transitions from $C=1$ to $C=0$. Here, $|1(2),\uparrow(\downarrow)\rangle$ denote the Dirac-cone states D1 and D2, where $\uparrow(\downarrow)$ indicate the spin state. $|\pm,\uparrow(\downarrow)\rangle$ denote the bonding ($+$) and antibonding ($-$) states of $|1(2),\uparrow(\downarrow)\rangle$. For the condition of $\Delta_{\textrm{intra}}=0$ and $m \neq 0$, both  D1 and D2 open an energy gap and contribute the Chern number $C_{1,2}=1/2$ for a total $C=C_1+C_2=1$. A finite intra-layer coupling $\Delta_{\textrm{intra}}$ couples the D1 and D2 to form bonding and antibonding states $|\pm,\uparrow(\downarrow)\rangle$. If we gradually increase $\Delta_{\textrm{intra}}$ from 0,  when $m>\Delta_{\textrm{intra}}$, the total Chern number preserves $C=1$; when $m=\Delta_{\textrm{intra}}$, the energy gap is closed; when $m<\Delta_{\textrm{intra}}$, the total Chern number become to be $C=0$. (c) A FM bi-BL multilayer with four Dirac-cone states (D1, D2, D3, D4). $\Delta_{\textrm{inter}}$ denotes the inter-layer coupling. (d) Schematic band structure of a FM bi-BL multilayer when $m=0$. V1,2 denote the valence bands, and C1,2 denote the conduction bands. If the magnetic exchange field is gradually induced, the V1,2 and C1,2 bands will split and lead to two topological phase transitions from $C=0$ to $C=1$ at $m=m_1$, then to $C=2$ at $m=m_2$. (e) The phase diagram on the parameters ($m/\Delta_{\textrm{inter}}$, $\Delta_{\textrm{intra}}/\Delta_{\textrm{inter}}$) of the FM bi-BL multilayer, calculated by the Dirac-fermion model. (f) A FM 5-BL multilayer and its phase diagram by the Dirac-fermion model. (g) Schematic magnetic multilayers consist of magnetic TIs.}
    \label{fig2}
\end{figure*}

\subsection{A. A single BL with FM magnetization}
We first consider a BL with a FM magnetization. For simplicity, the out-of-plane magnetization is taken as an example. The FM exchange field would induce Zeeman-type spin splitting through adding a mass term $ m\sigma_{z}$ to Eq. (\ref{eqBL}), which reads,
\begin{equation}\label{eqmBL}
    H_{\textrm{BL}}^{mm}(\textbf{k})=v(k_x \sigma_y-k_y \sigma_x )\tau_{z}+ \Delta_{\textrm{intra}}\tau_{x}+ m\sigma_{z},
\end{equation}
where the two superscripts $m$ and $m$ denote the FM exchange fields for the top and bottom Dirac-cone surface states, respectively, which are identical here by considering a uniform FM order. In addition, the sign of $m$ is determined by the FM order in the $+z$ or $-z$ direction. For simplicity, unless otherwise specified, $m$ is set to be positive with the FM order in the $+z$ direction. 

For the topologically nontrivial BL ($M<0$), there are two local Dirac-cone states (D1 and D2) on the top and bottom surfaces, as schematically shown in Fig.~\ref{fig2}(a). The competition of the intra-layer coupling $\Delta_{\textrm{intra}}$ and the FM exchange field $m$ can drive the two Dirac-cone surface states into different topological phases. In order to have a clear physical picture, we start from the limit of  $\Delta_{\textrm{intra}}=0$, where D1 and D2 are decoupled. Under the FM exchange field $m$, both D1 and D2 will open an energy gap, and each Dirac cone contributes a half Chern number ($C_{1,2}=1/2$) to the total Chern number ($C=C_1+C_2=1$) of the BL, and their bands are doubly degenerate, shown in Fig.~\ref{fig2}(b). Then we introduce the intra-layer coupling $\Delta_{\textrm{intra}}$ to lead to the bonding and anti-bonding states of D1 and D2, so the degeneracy of the bands of D1 and D2 is lifted, but the Chern number of the BL remains unchanged as $C=1$ as long as the energy gap keeps open for a small $\Delta_{\textrm{intra}}$. When further increasing the $\Delta_{\textrm{intra}}$, the energy gap is expected to close at a critical value of the $\Delta_{\textrm{intra}}$ and reopen for a larger $\Delta_{\textrm{intra}}$.  The gap closing-and-reopening process is accompanied by a topological phase transition with the Chern number changing from $C=1$ to $C=0$, shown in Fig.~\ref{fig2}(b). Such a process can be confirmed by the calculation based on the Dirac-fermion model in Eq.~(\ref{eqmBL}). The energy spectra of the four bands are given by $E_{\bf k}=\pm \sqrt{k^{2}+\left(m\mp \Delta_{\textrm{intra}}\right)^2}$, where $k=\sqrt{k_{x}^2+k_{y}^2}$. The Fermi level is located between the second and the third bands. The band gap at $\Gamma$ is $2\left|m-\Delta_{\textrm{intra}}\right|$ and the Chern number is obtained as
\begin{equation}\label{Chern}
C = \frac{1}{2}\left[ {{\text{sgn}}\left( {m + {\Delta_{\textrm{intra}}}} \right) + {\text{sgn}}\left( {m - {\Delta_{\textrm{intra}}}} \right)} \right].
\end{equation}
This confirms that the Chern number is $C=1$ for $\Delta_{\textrm{intra}}<m$, and a topological phase transition occurs at $\Delta_{\textrm{intra}}=m$ with a closed band gap, and the Chern number becomes $C=0$ for $ \Delta_{\textrm{intra}}>m$. Therefore, we can conclude that the competition between the FM exchange field and the intra-layer coupling induces a topological transition for a single FM BL. 

\subsection{B. A bi-BL structure with FM magnetization}
Then, we investigate a FM bi-BL structure to see the effect of the inter-layer coupling $\Delta_{\textrm{inter}}$, as schematically in Fig.~\ref{fig2}(c). With respect to Eqs.~(\ref{eqmultilayer},\ref{eqmBL}), the bi-BL Hamiltonian can be directly written as 
\begin{equation}
\label{FM-bi}
    H_{\textrm{FM}}^{bi}(\textbf{k})=\begin{pmatrix}
    H_{\textrm{BL}}^{mm}(\textbf{k})  &\tilde{\Delta}_{\textrm{inter}} \\
    \tilde{\Delta}_{\textrm{inter}}^{\dagger} & H_{\textrm{BL}}^{mm}(\textbf{k})
        \end{pmatrix}.
\end{equation}
There are four local Dirac-cone states, namely, D1 and D2 on the upper BL, D3 and D4 on the lower BL, seen in Fig.~\ref{fig2}(c). In the first stage, we turn on the intra-layer coupling $\Delta_{\textrm{intra}}$ and the inter-layer coupling $\Delta_{\textrm{inter}}$ in sequence but tentatively suppress the FM exchange field (e.g. $m=0$) for clarity. The bands of the four Dirac-cone states (D1-D4) should form four doubly degenerate bands (valence bands V1 and V2, conduction bands C1 and C2), shown in Fig.~\ref{fig2}(d), because of the time-reversal symmetry and the implicit inversion symmetry in this model. At this stage, the Chern number has to be $C=0$ due to the preserved the time-reversal symmetry in the absence of the FM exchange field. Then, we turn on the FM exchange field, and the double degeneracy of the bands will be gradually lifted. As the FM exchange field increases, two band inversions sequentially happen between the conduction and valence bands, corresponding to two topological phase transitions of ($C=0$ to $C=1$) and ($C=1$ to $C=2$), seen in Fig.~\ref{fig2}(d). The first topological phase transition happens along with the gap closing-and-reopening process between one V1 band and one C1 band.  By diagonalizing the Hamiltonian of Eq.~(\ref{FM-bi}), the corresponding band gap can be calculated as $\left|2m-\left(\bar{\Delta}-\Delta_{\textrm{inter}}\right)\right|$, where $\bar{\Delta}=\sqrt{4\Delta_{\textrm{intra}}^{2}+\Delta_{\textrm{inter}}^2}$. When the FM exchange field strength reaches $m_1=\left(\bar{\Delta}-\Delta_{\textrm{inter}}\right)/2$, the bi-BL structure undergoes a topological phase transition from $C=0$ to $C=1$. With further increasing the FM exchange field, the second topological phase transition occurs at the gap closing point between one V2 band and one C2 band, at the critical value $m_2=\left(\bar{\Delta}+\Delta_{\textrm{inter}}\right)/2$, corresponding to a change of the Chern number from $C=1$ to $C=2$. 

The detailed phase diagram, depending on the parameters ($m/\Delta_{\textrm{inter}}$, $\Delta_{\textrm{intra}}/\Delta_{\textrm{inter}}$), can be obtained from the Dirac-fermion model in Eq.~(\ref{FM-bi}) and is shown in Fig.~\ref{fig2}(e). In the limiting case of $\Delta_{\textrm{intra}}\rightarrow 0$,  the two Dirac-cone states (D1,2 or D3,4) of each BL are completely decoupled, so two Dirac-cone states (D1 and D4) of the bi-BL structure become isolated, and the other two Dirac-cone states (D2 and D3) couple together through the inter-layer coupling $\Delta_{\textrm{inter}}$. The isolated D1 and D4 contribute a Chern number ($C=1$), and the contributed Chern number of the coupled D2 and D3 depends on the competition of $m$ and $\Delta_{\textrm{inter}}$, for example, the Chern number $C=1$ for $m>\Delta_{\textrm{inter}}$, and $C=0$ for $m<\Delta_{\textrm{inter}}$, similar to the consideration in a FM BL. Therefore, based on the Dirac-fermion model, the highest Chern number of a FM bi-BL structure can reach but not exceed $|C|=2$. In order to obtain a higher Chern number, we should stack more FM BLs together. Straightforwardly, at the limit $\Delta_{\textrm{intra}}\rightarrow 0$, the Chern number of a FM $N$-BL structure is directly obtained as,
\begin{equation}
    C_N=1+\frac{N-1}{2}[\text{sgn}(m+\Delta_{\textrm{inter}} )+\text{sgn}(m-\Delta_{\textrm{inter}} )], 
\end{equation}
where the first constant $1$ comes from the two isolated Dirac-cone states on the top and bottom surfaces of the FM $N$-BL structure, and the second term is contributed by ($N-1$) pairs of coupled Dirac-cone states between adjacent BLs. The highest Chern number can be realized as $C=N$ when and only when $m > \Delta_{\textrm{inter}}$. As an example, five FM BLs are stacked together to the the highest Chern number $C=5$ at the limit $\Delta_{\textrm{intra}}\rightarrow 0$, seen in Fig.~\ref{fig2}(f). Therefore, we should stack at least $N$ FM BLs to realize a Chern number $C=N$ ($N \geq 1$), but we do not need to care for its details of each BL, within the framework of the Dirac-fermion approach. For example, we know that only the Chern number $C=1$ is realized in a Cr-doped (Bi,Sb)$_2$Te$_3$ 5-QL film in experiments~\cite{changcuizu2013experimental}, which can be effectively considered as a FM BL, though it consists of 5-QL Cr-doped (Bi,Sb)$_2$Te$_3$. 

\subsection{C. Magnetic TI multilayers}
Based on the Dirac-fermion approach, we now consider constructing magnetic TI multilayers through stacking magnetic/non-magnetic TI layers, for example, alternate stacking of a FM TI layer and a non-magnetic TI layer ($M < 0$), shown in Fig.~\ref{fig2}(g). The FM TI layer can be described by a FM $N$-BL structure, where $N$ is determined by the highest Chern number the FM TI layer can achieve, and the non-magnetic TI layer is described by a non-magnetic BL. In this sense, FM MnBi$_4$Te$_7$ can be effectively considered as a FM TI multilayer of a FM MnBi$_2$Te$_4$ BL and a non-magnetic Bi$_2$Te$_3$ BL. A FM TI multilayer can also consist of a FM TI layer and a non-magnetic trivial insulator layer, where the trivial insulator layer acts as a buffer layer to decrease the inter-layer coupling $\Delta_{\textrm{inter}}$. Besides the FM TI multilayers, AFM TI multilayers also can be considered. For example, an AFM TI multilayer consists of two FM TI layers with the opposite magnetization direction, shown in Fig.~\ref{fig2}(g). With the same consideration, each FM TI layer is described by a FM $N$-BL structure and $N$ is determined by the highest Chern number of the FM TI layer. For example, AFM MnBi$_2$Te$_4$ can be effectively considered as an AFM TI multilayer, in which one SL is considered as a FM BL.

\section{FM $\text{MnBi}_{2}\text{Te}_{4}$ films\label{sec4}}

\begin{figure}[tbp]
    \includegraphics[width=0.48\textwidth]{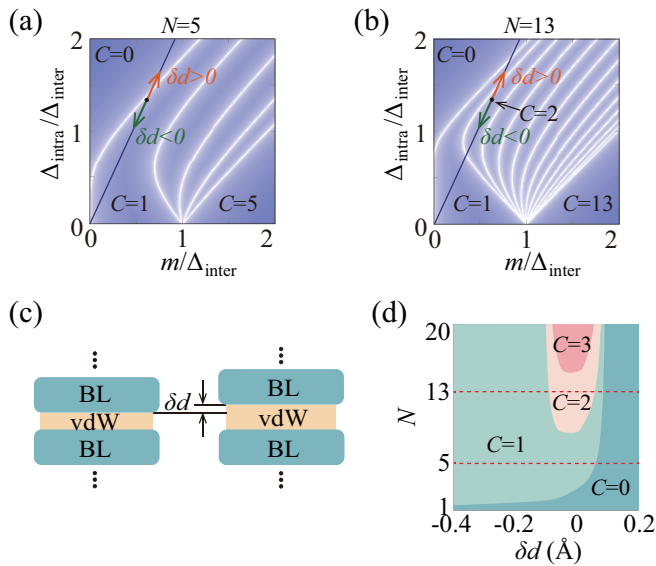}
    \caption{(a,b) The phase diagrams of FM 5-BL and 13-BL multilayers. The black points denote the phase of MnBi$_{2}$Te$_{4}$ without the vdW gap modulation. The red lines denote the phase evolution due to the vdW gap modulation with fixed $\Delta_{\textrm{intra}}$ and $m$. $\delta d>0$ ($\delta d<0$) indicates the vdW gap expansion (contraction). (c) The schematic of the vdW gap modulation ($\delta d$). (d) The phase diagram on the parameters (the number of SL $N$, the vdW gap modulation $\delta$) of AFM MnBi$_{2}$Te$_{4}$ films is calculated by the Dirac-fermion model with $m=40$ meV, $\Delta_{\textrm{intra}}=86.2$ meV, and $\Delta_{\textrm{inter}}=74.8$ meV from fitting first-principles calculations. The two red dashed lines in (d) correspond to the phase evolutions (black solid lines) in (a, b).}
    \label{fig3}
\end{figure}

In this section, we revisit the realization of the Chern insulator state with a high Chern number in FM MnBi$_{2}$Te$_{4}$ films. MnBi$_{2}$Te$_{4}$ is an intrinsic magnetic TI which provides an ideal platform to study the interplay between magnetism and topology \cite{gong2019cpl,zhangdongqin2019topological,lijiaheng2019intrinsic, otrokovmm2019unique, Sun2019Rational, gu2021spectral, otrokovPredictionObservationAntiferromagnetic2019,wang2020dynamical}. MnBi$_{2}$Te$_{4}$ has a layered crystal structure stacked by (Te-Bi-Te-Mn-Te-Bi-Te) SLs with a weak vdW interaction. Its magnetic ground state is an A-type AFM order with an out-of-plane magnetic anisotropy along the (111) direction. Although the AFM order breaks the time-reversal symmetry, the combined symmetry $S=T\tau_{1/2}$ of the time reversal operation ($T$) and a half-translation operation ($\tau_{1/2}$) is preserved in MnBi$_{2}$Te$_{4}$, which plays the role of the time-reversal symmetry and leads to a $Z_{2}$ topological classification with $M<0$ in the $k\cdot p$ Hamiltonian \cite{afmz2, zhangdongqin2019topological}. Interestingly, a moderate external magnetic field ($\sim 10$ T) can flop spins of the AFM phase to a FM phase MnBi$_{2}$Te$_{4}$ \cite{dengyujun2020quantum,gejun2020high,ying2022}. The FM  MnBi$_{2}$Te$_{4}$ was predicted to be an ideal Weyl semimetal in the bulk \cite{zhangdongqin2019topological,lijiaheng2019intrinsic}, which indicates that a Chern insulator state with a high Chern number should be realized in FM MnBi$_{2}$Te$_{4}$ films. 

Next, we investigate the Chern insulator state in FM phase MnBi$_{2}$Te$_{4}$ films through employing the Dirac-fermion approach. First of all, if a MnBi$_{2}$Te$_{4}$ SL is considered as a BL with $M<0$, then one FM MnBi$_{2}$Te$_{4}$ $N$-SL film can be seen as a FM $N$-BL TI multilayer. Secondly, the parameter of $\Delta_{\textrm{intra}}$ denotes the intra-layer coupling of the two Dirac-cone states of each MnBi$_{2}$Te$_{4}$ SL, the parameter of $\Delta_{\textrm{inter}}$ denotes the inter-layer coupling of the two Dirac-cone states between the adjacent MnBi$_{2}$Te$_{4}$ SLs, and the parameter $m$ denotes the FM exchange field within each MnBi$_{2}$Te$_{4}$ SL.

We then take the FM $5$-SL MnBi$_{2}$Te$_{4}$ film as an example. The Dirac-fermion model can be directly set up for the FM $5$-BL TI multilayer through Eq.~(\ref{eqmultilayer}), and a phase diagram can be obtained on the parameters ($\Delta_{\textrm{intra}}/\Delta_{\textrm{inter}}$, $m/\Delta_{\textrm{inter}}$), shown in Fig.~\ref{fig3}(a). Though the Chern numbers from 0 to 5 are expected to be realized through tuning $\Delta_{\textrm{intra}}/\Delta_{\textrm{inter}}$ and $m/\Delta_{\textrm{inter}}$, the realistic Chern number of the FM $5$-SL MnBi$_{2}$Te$_{4}$ film is calculated to be $C=1$ from material-dependent parameters ($\Delta_{\textrm{intra}}$, $\Delta_{\textrm{inter}}$ and $m$). Moreover, we have also calculated the phase diagram for the FM $13$-BL TI multilayer, shown in Fig.~\ref{fig3}(b), where we can see that the highest Chern number $C=13$ can be obtained at the limiting situation of ($\Delta_{\textrm{intra}}/\Delta_{\textrm{inter}}\rightarrow 0$, $m/\Delta_{\textrm{inter}}>1$), while the realistic Chern number of the FM phase $13$-SL MnBi$_{2}$Te$_{4}$ film is calculated to be $C=2$. On the whole, the results of FM phase $5$- and $13$-SL MnBi$_{2}$Te$_{4}$ films are consistent with the reported experiments~\cite{dengyujun2020quantum,gejun2020high,ying2022}.   

Further, we consider the effect of the vdW gap modulation ($d_0+\delta d$) of FM MnBi$_{2}$Te$_{4}$ films through external pressures, where $d_0$ denotes the pristine vdW gap. The in-plane pressure is expected to induce a vdW gap expansion ($\delta d>0$), and the out-of-plane pressure is to induce a vdW gap contraction ($\delta d<0$). We assume the inter-layer coupling $\Delta_{\textrm{inter}}$ changes exponentially to the deviation ($\delta d$) from the pristine vdW gap, $\Delta_{\textrm{inter}}(\delta d)=\Delta_{\textrm{inter}}\text{exp}{(-\lambda \delta d)}$ where $\lambda$ denotes the changing rate. The effect of the vdW gap modulation in the phase diagram of FM $5$- and $13$-SL MnBi$_{2}$Te$_{4}$ films is shown in Figs.~\ref{fig3}(a) and \ref{fig3}(b), respectively. For the FM $5$-SL MnBi$_{2}$Te$_{4}$ film, we can see that the vdW gap expansion ($\delta d>0$) can induce a topological phase transition from $C=1$ to $C=0$, and that the vdW gap contraction ($\delta d<0$) prefers to keep the ($C=1$) phase. But it is quantitatively different for the FM $13$-SL MnBi$_{2}$Te$_{4}$ film. The vdW gap expansion ($\delta d>0$) can induce two topological phase transitions from $C=2$ to $C=1$ and then $C=1$ to $C=0$, and the vdW gap contraction ($\delta d<0$) induces a topological phase transition from $C=2$ to $C=1$. Based on the Dirac-fermion model, we carry out systematic calculations on the pressure-induced phase diagrams for FM $N$-SL MnBi$_{2}$Te$_{4}$ films with $N$ from 1 to 20, shown in Fig.~\ref{fig3}(d). We can notice that a large vdW gap expansion ($\delta d>0$) always leads to a topologically trivial phase with $C=0$, which is well understood because it is at the decoupled limit situation. Interestingly, a large vdW gap contraction ($\delta d<0$) also prefers to induce a low Chern number $C=1$, not a high Chern number. The high Chern number only arises for a moderate vdW gap modulation, for example, $C=2$ for $N=13$ and $C=3$ for $N=20$, which is an important consequence from the Dirac-fermion approach.

\section{nonmagnetic TI/magnetic insulator multilayers \label{sec5}}

%Besides the intrinsic magnetic topological insulator, early reseaches introduced magentic dopant to topological insulators (Bi,Sb)$_{2}$Te$_{3}$ films which can also realize QAH~\cite{zhangjinsong2013topology,changcuizu2013experimental, kouxufeng2014scale, CheckelskyNP2014,xiaodiPRL2018, grauerPRL2017,MasatakaSciAdv2017,BestwickPRL2015}.

Recently, Zhao \textit{et al.} \cite{zhaoyifanPRL2022, zhaoyifan2020tuning} fabricated magnetic TI multilayers consisting of alternate [3QL-(Bi,Sb)$_{1.76}$Cr$_{0.24}$Te$_{3}$ (BCT)]/[4QL-(Bi,Sb)$_{2}$Te$_{3}$ (BT)] bilayers. Notably, a high Chern number of $C=5$ has been experimentally realized in these multilayers \cite{zhaoyifanPRL2022}. Moreover, a topological phase transition from $C=2$ to $C=1$ has been observed in a [3QL-BCT/4QL-BT/3QL-BCT/4QL-BT/3QL-BCT] multilayer through changing Cr doping of the middle 3QL-BCT  layer \cite{zhaoyifan2020tuning}.  In the following, we will present a understanding of the two phenomena in these multilayers from the viewpoint of the Dirac-fermion approach.

\subsection{A. A high Chern number $C=N$ case}

\begin{figure}[tbp]
    \includegraphics[width=0.45\textwidth]{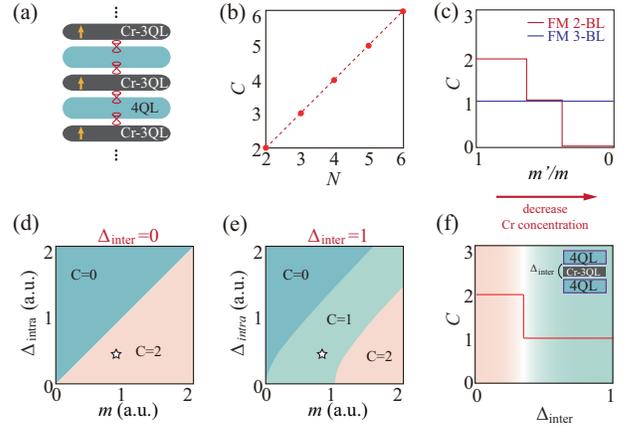}
    \caption{(a) The schematic structure of [3QL-BCT]/[4QL-BT] multilayers. (b) The linearly increased Chern number with the number of [3QL-BCT/4QL-BT] layers, in the case of weak inter-layer and intra-layer couplings. (c) Chern number evolution from the 2-BL (red line) and 3-BL (blue line) Dirac model due to decreasing the Cr doping in the middle 3QL-BCT layer. We take the parameters  $m=0.7$, $\Delta_{\textrm{intra}}=0.2$, and $\Delta_{\textrm{inter}}=0.1$ for the 2-BL model, and $\Delta_{\textrm{inter}}=1$, and $m_{0}=2m^{\prime}$ for the 3-BL model. (d,e,f) Phase diagram for the 2-BL model with $m^{\prime}=m$, and the inter-layer coupling $\Delta_{\textrm{inter}}=0$ (d) and $\Delta_{\textrm{inter}}=1$ (e). A finite $\Delta_{\textrm{inter}}$ induces a $C=1$ phase (e). Topological phase transition from $C=2$ to $C=1$ arises when the inter-layer coupling $\Delta_{\textrm{inter}}$ is continuously increased from 0 to 1 (f). }
    \label{fig4}
\end{figure}

For a TI multilayer structure constructed by $N$ alternating [3QL-BCT]/[4QL-BT] layers, shown in Fig.~\ref{fig4}(a), each 4QL-BT layer can be chosen as a nonmagnetic [4QL-BT] BL. Two Dirac-cone states on the top and bottom surfaces of each [4QL-BT] BL are weakly coupled through a weak intra-layer coupling $\Delta_{\textrm{intra}}$ because of the thickness of [4QL-BT] BL. Differently, because the spin-orbit coupling (SOC) of Cr atom is negligible, the effective SOC in the 3QL-BCT layer is expected to be much reduced due to the heavily doped Cr through replacing Bi by Cr, so the bulk BCT becomes a topologically trivial FM insulator. On one side, the presence of the 3QL-BCT layer significantly reduces the inter-layer coupling $\Delta_{\textrm{inter}}$ between neighboring [4QL-BT] BLs, which enables us to take the limit of $\Delta_{\textrm{inter}}\rightarrow0$ to have a clear physical picture. On the other side, the 3QL-BCT layer provides a FM proximity effect to its adjacent [4QL-BT] BLs, which brings a magnetic exchange field $m$ to the Dirac-cone states of the [4QL-BT] BL. As discussed in Sec.~\ref{sec3}, for each [4QL-BT] BL, the competition between $\Delta_{\textrm{intra}}$ and $m$ should lead to different topological phases, namely, $C=1$ when $m>\Delta_{\textrm{intra}}$, and $C=0$ when $m<\Delta_{\textrm{intra}}$.

In this decoupled limit, the total Hamiltonian from the Dirac-fermion model can be simply written as a direct sum of $N$ [4QL-BT] BLs 
\begin{equation}\label{H1}
    H_{N}(\textbf{k})=\bigoplus_{i=1}^{N}H_{i,\textrm{BL}}^{mm}(\textbf{k}),
\end{equation}
where $i$ is the [4QL-BT] BL index and $m$ denotes the magnetic exchange field on each Dirac-cone state. According to the discussion in Sec.~\ref{sec3}, the total Chern number of the multilayer is then given by 
\begin{equation}\label{CN}
    C=\frac{N}{2}\left[\text{sgn}\left(m+\Delta_{\textrm{intra}}\right)+\text{sgn}\left(m-\Delta_{\textrm{intra}}\right)\right]. 
\end{equation}
Therefore,  if $m > \Delta_{\textrm{intra}}$, a high Chern number of $C=N$ can be obtained in $N$-[3QL-BCT]/[4QL-BT] multilayers, as illustrated in Fig.~\ref{fig4}(b), which is consistent with the experimental observations \cite{zhaoyifan2020tuning}.

\subsection{B. Topological phase transition from $C=2$ to $C=1$}

Then, we focus on the [3QL-BCT/4QL-BT/3QL-BCT/4QL-BT/3QL-BCT] multilayer to understand the experimentally observed topological phase transitions from $C=2$ to $C=1$ induced by reducing the Cr doping \cite{zhaoyifanPRL2022} of the middle 3QL-BCT layer. Generally speaking, there are two effects if the Cr doping amount is decreased from the heavily doping limit. First of all, the FM exchange field produced by moments of Cr atoms should be accordingly decreased. Secondly, the SOC in the 3QL-BCT layer should be enhanced, and the energy gap of the 3QL-BCT layer is accordingly reduced, which would lead to an enlarged effective inter-layer coupling $\Delta_{\textrm{inter}}$ between two neighboring [4QL-BT] BLs. It should also be noted that with decreasing the Cr doping amount, the BCT could possibly becomes topologically non-trivial with a band inversion ($M<0$) hosting extra Dirac-cone states. In the following, depending on whether the middle 3QL-BCT layer is topologically trivial or not, we will set up a 2-BL or 3-BL model to discuss the phase transition from $C=2$ to $C=1$.

Firstly, when the middle 3QL-BCT layer is topologically trivial,  the whole multilayer system can be effectively described by the Dirac-fermion model consisting of two [4QL-BT] BLs, with the Hamiltonian given by 
\begin{equation}\label{FM 2-BL}
    H_{2\textrm{BL}}(\textbf{k})=\begin{pmatrix}
    H_{\textrm{BL}}^{mm^{\prime}}(\textbf{k}) &  \tilde{\Delta}_{\textrm{inter}}  \\
      \tilde{\Delta}_{\textrm{inter}} ^{\dagger}& H_{\textrm{BL}}^{m^{\prime}m}(\textbf{k})
        \end{pmatrix},
\end{equation}
where 
\begin{equation}\label{BLmmp}
    H_{\textrm{BL}}^{mm^{\prime}}=\begin{pmatrix}
    H_{D}^{m}(\textbf{k}) & \Delta_{\textrm{intra}} \\
    \Delta_{\textrm{intra}} & H_{D}^{m^{\prime}}(\textbf{k})
        \end{pmatrix}.
\end{equation}
Here, $H_{\textrm{BL}}^{mm^\prime}$ indicates the Dirac-fermion model for a [4QL-BT] BL, and $m$ ($m^{\prime}$) denotes the magnetic proximity exchange filed acting on top (bottom) Dirac-cone states, respectively. For simplicity, we first consider the case with $\Delta_{\textrm{inter}}=0$, where the two BLs are decoupled, which enables us to independently consider $H_{\textrm{BL}}^{mm^\prime}$ or $H_{\textrm{BL}}^{m^{\prime}m}$. Since only the Cr-doping of the middle 3QL-BCT layer is decreased, $m'<m$ should be expected. Under this condition, it can be shown that, for each [4QL-BT] BL, when $m^{\prime}>\Delta_{\textrm{intra}}^2/m$ ($m^{\prime}<\Delta_{\textrm{intra}}^2/m$) is satisfied, the Chern number is obtained as $C_{\textrm{BL}}=1$ ($C_{\textrm{BL}}=0$). Therefore, when decreasing $m'$ from $m$ ($m>\Delta_{\textrm{intra}}$  is assumed) along with the gradually reduced Cr-doping, the multilayer should have a topological phase transition directly from $C=1+1=2$ to $C=0+0=0$, without an intermediate $C=1$ phase, shown in Fig.~\ref{fig4}(c). The phase diagram on parameters ($\Delta_{\textrm{intra}}$, $m$) is calculated and shown in Fig.~\ref{fig4}(d), and we also see that there are only two regions of $C=2$ and $C=0$. Therefore, though the $C=2$ Chen insulator state is well captured, the results do not reproduce the topological phase transition from $C=2$ to $C=1$ observed in experiments \cite{zhaoyifanPRL2022}. Then, we will show that the missing $C=1$ phase emerges when taking the inter-layer coupling $\Delta_{\textrm{inter}}$ into account. For simplicity, we fix $m^\prime=m$. When we consider a finite inter-layer coupling (e.g. $\Delta_{\textrm{inter}}=1$), a $C=1$ phase region will emerge in the phase diagram between the $C=0$ and $C=2$ phases, as shown in Fig.~\ref{fig4}(e). This can be easily understood based on the 2-BL Dirac-fermion model, schematically shown in Fig.~\ref{fig2}(d). A finite inter-layer coupling pushes up the conduction bands C2 and pulls down the valence bands V2, and then causes a topological phase transition from $C=2$ to $C=1$, shown in Fig.~\ref{fig4}(f). Therefore, the inter-layer coupling $\Delta_{\textrm{inter}}$ is indispensable for the phase transition from $C=2$ to $C=1$, and the enhanced inter-layer coupling with decreasing the Cr doping further facilitates the experimental observation of this transition in [3QL-BCT/4QL-BT/3QL-BCT/4QL-BT/3QL-BCT] multilayer \cite{zhaoyifanPRL2022}.

Secondly, when the middle 3QL-BCT layer becomes topologically nontrivial in the low Cr-doping regime, we need to build a 3-BL Dirac-fermion model including two [4QL-BT] BLs and one [3QL-BCT] BL, and the Hamiltonian is written as, 
\begin{equation}\label{FM 3-BL}
    H_{3\textrm{BL}}(\textbf{k})=\begin{pmatrix}
    H_{\textrm{BL}}^{mm^{\prime}}(\textbf{k})  &\tilde{\Delta}_{\textrm{inter}}   &   \\
    \tilde{\Delta}_{\textrm{inter}}^{\dagger} & H_{\textrm{BL}}^{m_{0}m_{0}}(\textbf{k}) & \tilde{\Delta}_{\textrm{inter}}\\
    & \tilde{\Delta}_{\textrm{inter}}^{\dagger} & H_{\textrm{BL}}^{m^{\prime}m} (\textbf{k}) 
    \end{pmatrix},
\end{equation}
where $m_{0}$ is the magnetic exchange filed acting on the Dirac states of the middle [3QL-BCT] BL. Because the inter-layer coupling $\Delta_{\textrm{inter}}$ is much enhanced with decreased Cr doping, the three BLs are strongly bound together, compared with the 2-BL Dirac-fermion model. It is found that within a large parameter region, the whole multilayer system retains $C=1$, seen in Fig.~\ref{fig4}(c).

We should stress that the 3-BL Dirac-fermion model is valid only for the topologically nontrivial middle 3QL-BCT layer, while the 2-BL Dirac-fermion model is applicable for the topologically trivial middle 3QL-BCT layer. Therefore, if the BCT changes from a topologically trivial phase ($M>0$) to a topologically nontrivial phase ($M<0$) with decreasing the Cr doping, the 3-BL and 2-BL Dirac-fermion model should be combined to describe the observed topological phase transition from $C=2$ to $C=1$ for the [3QL-BCT/4QL-BT/3QL-BCT/4QL-BT/3QL-BCT] multilayer in the experiments \cite{zhaoyifanPRL2022}.

\section{Summary\label{sec6}}

To summarize, we build a Dirac-fermion approach to effectively describe the low-energy electronic structure of various multilayers of layered (magnetic) TIs. Based on the Dirac-fermion model, we demonstrate how to design Chern insulator states with a high Chen number in multilayers of layered magnetic TIs. As an example, we employ the Dirac-fermion model to investigate the Chen insulator state of FM MnBi$_2$Te$_4$ films under the vdW gap modulation. We calculated the phase diagram and found that a moderate vdW gap modulation is necessary to realize a high Chen number for FM MnBi$_2$Te$_4$ thick films. We also employ the Dirac-fermion approach to understand the topological phase transition in the [3QL-BCT/4QL-BT] multilayers. Weak intra-layer and inter-layer couplings and a strong magnetic exchange proximate field are the key to realizing the high Chern number of $C=N$ in $N$-[3QL-BCT]/[4QL-BT] multilayers. Further, the Dirac-fermion model reproduces the topological phase transition from $C=2$ to $C=1$ for  the [3QL-BCT/4QL-BT/3QL-BCT/4QL-BT/3QL-BCT] multilayer when the Cr doping is decreased, which reveals that either a topologically trivial middle 3QL-BCT layer with an enhanced inter-layer coupling or a topologically nontrivial middle 3QL-BCT layer can induce the topological phase transition. Therefore, the developed Dirac-fermion approach provides a powerful tool to understand the topological phase transition in various of multilayers of layered magnetic TIs. 

\begin{acknowledgements}
This work is supported by National Key Projects for Research and Development of China (Grant No.2021YFA1400400 and No.2017YFA0303203), the Fundamental Research Funds for the Central Universities (Grant No. 020414380185), Natural Science Foundation of Jiangsu Province (No. BK20200007), the Natural Science Foundation of China (No. 12074181, No. 12104217, and No. 11834006) and the Fok Ying-Tong Education Foundation of China (Grant No. 161006). Dinghui Wang is supported by the program A/B for Outstanding PhD candidate of Nanjing University.\\
\end{acknowledgements}

\appendix

\section{Surface states wavefunction}\label{appendix1}

For a thin film of a three-dimensional TI located within $|z|<L/2$ in the $z$-direction, we use the quantum-well-based methods~\cite{Liu2010Model}, where the basis of the quantum well states is given by
\begin{equation}
    \langle{z}\ket{n\alpha}=\sqrt{\frac{2}{L}}\sin\Big(\frac{z}{L}n\pi+\frac{n\pi}{2}\Big)\ket{\Lambda_{\alpha}}.
\end{equation}
where $n=1,2,3...$, and $\ket{\Lambda_{\alpha}}$ ($\alpha=1,2,3,4$) denotes the basis of the Hamiltonian of the TI in Eq. (\ref{eq1}). By the substitution $k_{z}\rightarrow -i\partial_{z}$ in the original four-band Hamiltonian, we can obtain $H_{nm}=\bra{n}H_{3D}(k_{x}, k_{y}, k_{z}\rightarrow-i\partial_{z})\ket{m}$, with the help of the relations
\begin{align}
    &\braket{n|m}=\delta_{nm},\\
    &\bra{n}-i\partial_{z}\ket{m}=\frac{2}{L}\frac{(-1+(-1)^{m+n})nm}{n^{2}-m^{2}},\\
    &\bra{n}-\partial^{2}_{z}\ket{m}=\left(\frac{n\pi}{L}\right)^{2}\delta_{nm}.
\end{align}

By choosing a cut-off number of $N$ of the quantum well states and diagonalizing the $4N\times 4N$ Hamiltonian $H_{nm}$, the $i$-th eigenstates of the quantum well is then obtained as
\begin{equation}
    \ket{\varphi^{i}(z)}=\sqrt{\frac{2}{L}}\sum _{\alpha=1}^{4}\sum_{n=1}^{N}c^{i}_{\alpha,n}\sin\Big(\frac{z}{L}n\pi+\frac{n\pi}{2}\Big)\ket{\Lambda_{\alpha}},
\end{equation}
where $c^{i}_{\alpha,n}$ is the superposition coefficient. Then the probabilitiy density for the wavefunction reads
\begin{equation}
    |\varphi^{i}(z)|^2=\frac{2}{L}\sum_{\alpha=1}^{4}\left|\sum_{n=1}^{N}c^{i}_{\alpha,n}\sin\Big(\frac{z}{L}n\pi+\frac{n\pi}{2}\Big)\right|^2.
\end{equation}
For the topologically nontrivial case, we focus on the bands located inside the bulk gap, while for topologically trivial case, the highest conduction bands are chosen. In the numerical calculations, we choose the same parameters as those in Ref. \cite{Liu2010Model}.

Following Eq. (2.32) in Ref.~\cite{shen2012topological}, we construct a exponentially decay function to decomposed the surface wave function, which reads
\begin{equation}
    f(z)=A\left(e^{\tilde{z}/\xi_{+}}-e^{\tilde{z}/\xi_{-}}\right),
\end{equation}
where $A$ is the normalized coeffecient, and $\xi_{\pm}^{-1}=\lambda_{0}-\lambda$. This function can well capture properties of density of the surface states. In a narrow quantum well, this function is used to decompose the combined surface states in Figs. \ref{fig1}(f)-(h).

\bibliography{ref}

\end{document}